\pdfoutput=1
\documentclass[aps,prl,twocolumn,groupedaddress]{revtex4}

\usepackage{graphicx}
\usepackage{amsmath}
\usepackage{color}

\begin{document}

\title{Strong gyrotopy in a chiral toroidal medium}

\author{N. Papasimakis}
\affiliation{Optoelectronics Research Centre, University of
Southampton, SO17 1BJ, UK} \email{N.Papasimakis@soton.ac.uk}

\author{V. A. Fedotov}
\affiliation{Optoelectronics Research Centre, University of
Southampton, SO17 1BJ, UK}

\author{K. Marinov}
\affiliation{Optoelectronics Research Centre, University of
Southampton, SO17 1BJ, UK}

\author{N. I. Zheludev}
\affiliation{Optoelectronics Research Centre, University of
Southampton, SO17 1BJ, UK}

\author{A. D. Boardman}
\affiliation{Institute for Materials Research, University of
Salford, M5 4WT, UK}

\date{\today}

\begin{abstract}
In this letter, we present the first experimental study of a new
chiral metamaterial consisting of toroidal wire windings. We show
that the metamaterial exhibits three bands of circular dichroism in
the GHz range. We discuss the response of the structure in terms of
multipole moments, including the (magnetic) toroidal dipole moment.
\end{abstract}

\maketitle


Toroidal, doughnut-shaped structures are ubiquitous in nature,
appearing on scales which range from the sub-atomic
\cite{zeldovich58, dubovik74} to the astronomical
\cite{mainstone67}. On the molecular level, the torus shape is
preferred by numerous biological and chemical macromolecules, such
as DNA condensates \cite{dnarev}, proteins \cite{protrev} and
oligosaccharides \cite{cyclod}, to name just a few. Toroidal
symmetries are also encountered frequently in solid-state systems
including carbon nanotubes \cite{carbon} and ferroelectrics
\cite{ferroelc, ferroel}. Although the behavior of such systems has
been studied extensively, their interaction with electromagnetic
radiation is not well understood. Indeed, within the framework of
classical electrodynamics, unusual phenomena associated with
violation of Lorentz reciprocity \cite{afanasiev01} and
non-radiating configurations \cite{afanasiev98} have been predicted
for toroidal structures and their interactions. Nevertheless, such
phenomena are usually weak \cite{sawada05, fedotov} and, therefore,
experimental investigations are rare. In this letter, we study a new
artificial chiral medium, originally suggested in \cite{tormetam},
consisting of an array of toroidal wire windings in a metamaterial`
configuration. We show that such a metamaterial exhibits strong
gyrotropic response which is attributed to different terms of its
multipole expansion.

In contrast to artificial gyrotropic media, where handedness is
usually associated with the direction of a "twist vector" following
a cork-screw law along the helicity axis, the situation is more
complicated when the structure possesses toroidal symmetry. Here,
the twist vector rotates along the torus, and therefore a
corresponding direction can not be defined. However, although no
helicity axis exists, the structure has two well-defined
enantiomeric forms, corresponding to different directions of the
winding (see Fig.~1a). Based on this concept, we manufactured a
chiral toroidal metamaterial with a unit cell consisting of four
connected square loops that were formed by horizontal and vertical
copper wire segments (see Fig.~1b). The resulting windings were
embedded in dielectric bars with permittivity $\epsilon=4.5-0.081i$.
The size of the unit cell was $15x15~mm$ rendering the metamaterial
non-diffracting up to $20~GHz$. Transmission experiments were
performed at normal incidence, from $2$ to $14~GHz$, in an anechoic
chamber using two broadband horn antennas (Schwarzbeck M. E. model
BBHA 9120D). The transmission spectra were recorded with a vector
network analyzer.

\begin{figure}[ht] \label{fig1}
\includegraphics[width=.5\textwidth]
{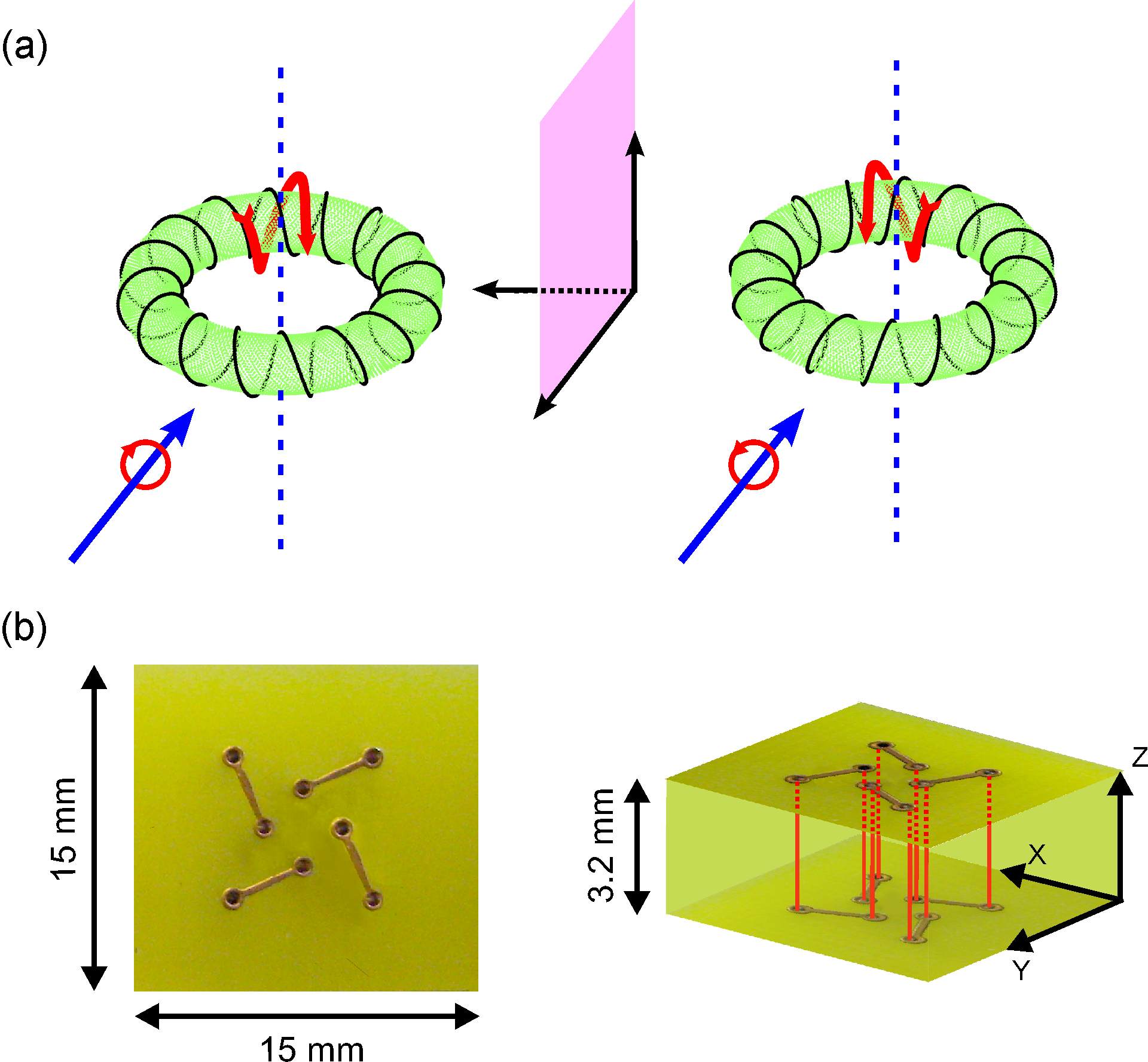}\caption{(a) Left- and right-handed form of a chiral
toroidal winding. The electromagnetic response of the structure is
determined by the handedness of the winding and that of the incident
wave. (b) Top and side view of the fabricated toroidal metamaterial
unit cell.}
\end{figure}

    The intensity and phase of the structure's response to right and
    left circularly polarized light are presented in Figs. 2a and
    2b, respectively. As it can be seen in Fig.~2a, experimental
    results (solid line) are in good agreement with finite element
    numerical simulations (solid circles). Two resonant bands of
    strong circular dichroism can be distinguished at around $4.5$
    and $10~GHz$, where the corresponding transmission ratio for
    orthogonal circularly polarized states reaches $5$ and $15~dB$
    with bandwidth (FWHM) $370~MHz$ and $1.6~GHz$, respectively.
    Moreover, while the rejected polarization state experiences high
    losses, the orthogonal polarization state propagates through the
    structure almost unaffected. This is also evident from the
    behavior of the phase delay presented in Fig. 2b. In addition,
    the rejected polarization state corresponds to a backward wave,
    while the other to a normal propagating wave, since they both
    have the same sign of phase velocity, but different signs of
    group velocity. This can be considered as a plausible indication
    for the presence of negative refraction \cite{pendry04}.
    Finally, a third band of circular dichroism can also be seen at
    around $7.5~GHz$. In this case, however, the dichroism is
    weaker, while both polarization states experience significant
    losses.

\begin{figure}[ht] \label{fig2}
\includegraphics[width=.5\textwidth]
{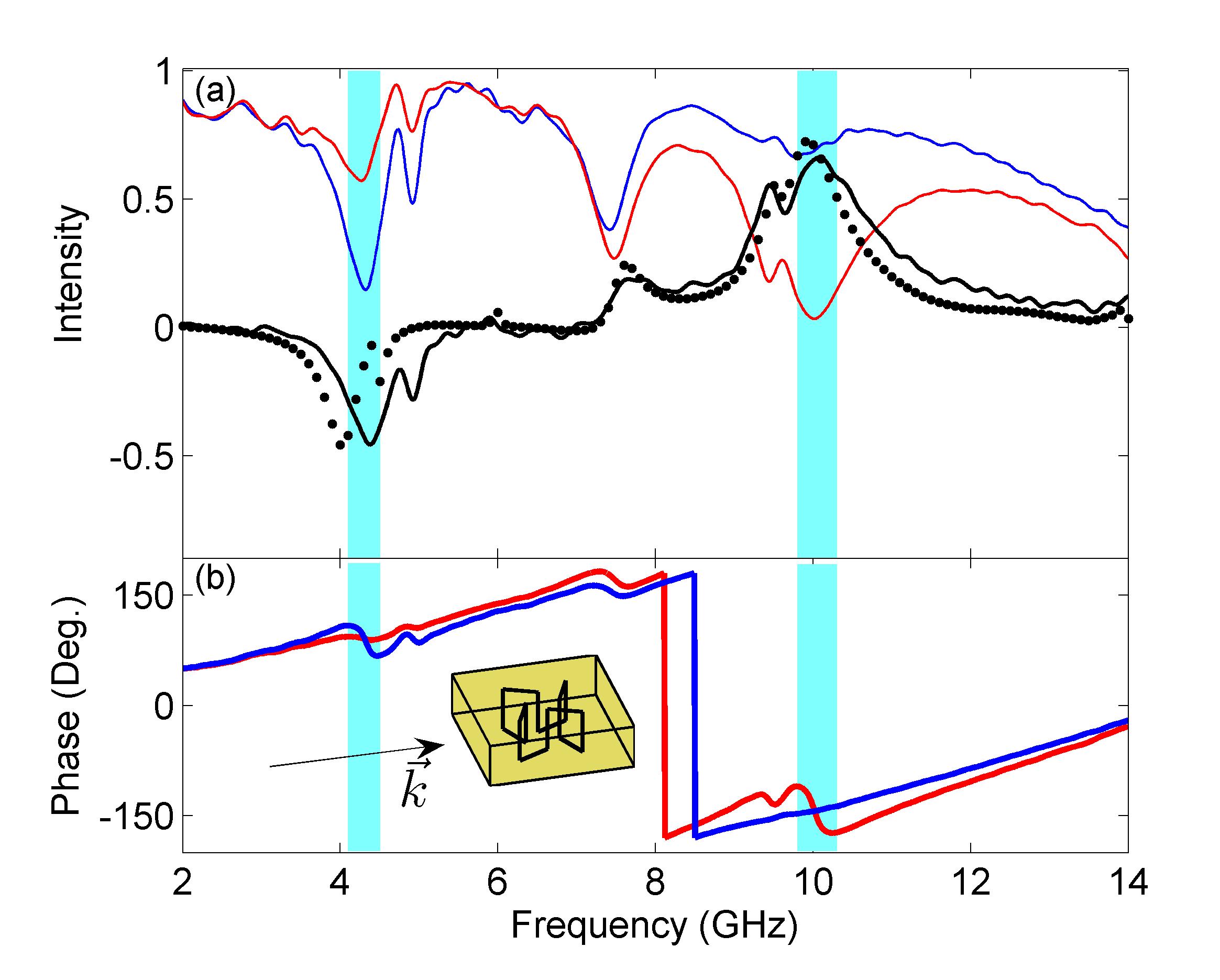}\caption{(a) Transmitted intensity for right (blue) and
left (red) circularly polarized light. The circular dichroism is
represented by the solid black curve (experimental) and the black
solid circles (numerical). The position of the dichroism resonances
is marked by the blue panels. (b) Phase delay for right (blue) and
left (red) circularly polarized light. Inset: Unit cell of the
studied metamaterial and direction of wave propagation.}
\end{figure}

The observed resonances are of geometrical origin and their
frequency position is controlled by the ratio of the total wire
length in the metamaterial unit cell over the wavelength of the
incident wave, while the presence of the dielectric has to be taken
also into account. In general, geometrical resonances will occur
whenever, the total wire length equals an integer number of
half-wavelengths. However, due to the symmetry of the structure only
an even number of nodes is allowed. Therefore, resonances will occur
whenever the wire length equals an integer multiple of the
excitation wavelength (even number of nodes). At these resonances,
the intensity of interaction is very different for left and right
circular polarization. This is illustrated in Figs.~3a \& 3b, where
energy density maps and power flow lines (following the Poynting
vector) are shown. While one polarization interacts strongly with
the structure (Fig.~3a) the orthogonal circular polarization
propagates almost unaffected. This behavior is also demonstrated in
Figs. 3b and 3c, where the real part of the current intensity along
the wire winding is plotted for the case of the low frequency
($4~GHz$) and high frequency ($10~GHz$) resonances. Although the
current configuration is very similar for both polarizations,
resembling standing waves with four ($4~GHz$) and six nodes
($10~GHz$), its intensity is significantly different which leads to
the observed dichroism resonances.

\begin{figure}[ht] \label{fig3}
\includegraphics[width=.5\textwidth]
{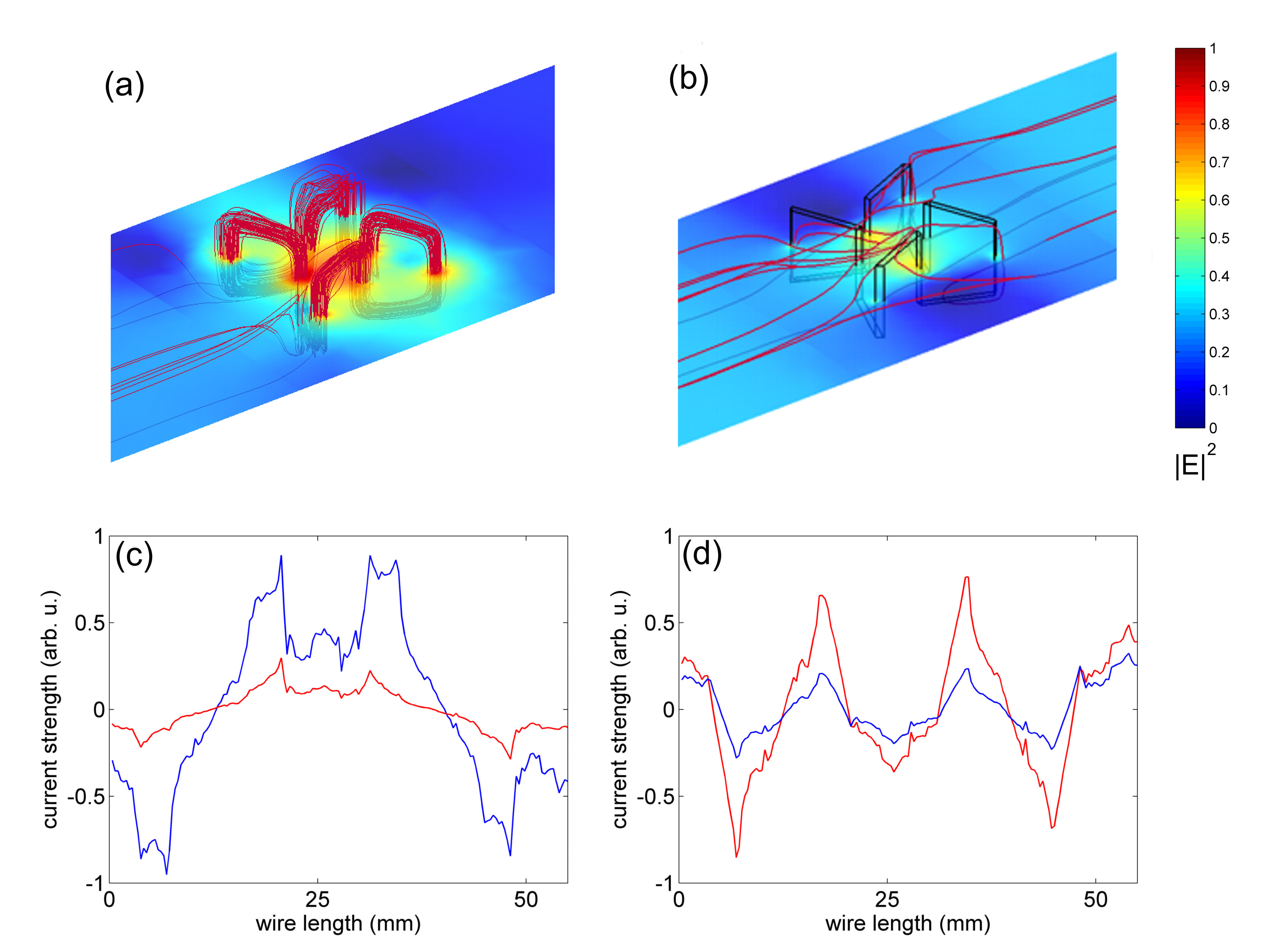}\caption{ Finite element simulations of toroidal
metamaterial. (a) and (b) show electric density (color maps) and
powerflow lines at the $4~GHz$ resonance for right and left
circularly polarized incident wave, respectively. (c) and (d)
present induced current configuration on the wire winding for right
(red) and left (blue) circularly polarized light, at the $4~GHz$ and
$10~GHz$ resonances, respectively.}
\end{figure}

    In the framework of multipole theory, the observed gyrotropy
    arises from the coupling of the electric and magnetic dipole
    moments, but can also include a contribution from the electric
    quadrupole moment, since the metamaterial is anisotropic
    \cite{buckingham71,theron96}. In order to further investigate
    the metamaterial's response, we calculate the induced multipole
    moments \cite{vrejoiu} under plane wave illumination
The multipole moments have been calculated by the following
relations:
    \begin{eqnarray*}
electric~dipole: & p_i=\frac{1}{i\omega}\int{J_idV} \\
magnetic~dipole: & m_i=\frac{1}{2}\int{(\vec{r}\times\vec{J})_idV}\\
electric~quadrupole: & Q_{ij}=\frac{1}{i\omega}\int{(r_iJ_j+J_ir_j)dV},\\
    \end{eqnarray*}
where $i,j=x,y,z$ and compared their strength in terms of radiated
power:

\begin{eqnarray*} el.~dipole~radiation: &
P_p=\frac{\omega^4\mu_0}{12\pi
c}|\vec{p}|^2\\
m.~dipole~radiation: & P_m=\frac{\omega^4\mu_0}{12\pi
c^3}|\vec{m}|^2\\
el.~quadrupole~radiation: & P_Q=\frac{\omega^6\mu_0}{160\pi
c^3}\sum|Q_{ij}|^2,\\
\end{eqnarray*}
 as presented in Fig. 4. As expected, the electric dipole moment dominates the
    response of the system at all frequencies, while all multipole
    moments resonate together at the $4~GHz$ and $10~GHz$ circular
    dichroism bands. More concisely, at the low frequency resonance, the magnetic dipole is
    much stronger than the electric quadrupole, indicating that the
    main contribution to chirality comes from the electric dipole-magnetic dipole
   coupling. This is also supported by the fact that these moments are almost collinear as shown in the
   corresponding inset of Fig.~4. Similar situation arises also at the
   high frequency dichroism band ($10~GHz$). However, at the
    weaker $7~GHz$ resonance, the electric quadrupole presents
    a pronounced resonance, while the magnetic dipole moment is
    weaker. This suggests that the weak dichroism at this resonance is
    a result of electric dipole - electric quadrupole coupling.
    Although ideally the electric
    dipole moment should also be at a minimum, the dominant contribution is still
    coming from the latter, as a result of the finite size of the structure.

    In addition, we include in our analysis the (magnetic) toroidal
    dipole moment \cite{zeldovich58}. Indeed, it has been suggested
    that the complete multipole expansion requires the inclusion of
    toroidal moments along with the electric and magnetic ones
    \cite{dubovik90}. While usually these moments can be neglected,
    this is not true for structures comparable to the
    wavelength or structures of toroidal symmetry \cite{dubovik90},
    as the one considered here. In fact, the toroidal moment is
    known to result in magneto-electric coupling \cite{fiebig05},
    since it enters the multipole expansion in exactly the same way
    as the electric moments \cite{dubovik90}. Therefore one can
    expect that coupling between toroidal and magnetic moments would
    lead to optical activity and the question of a toroidal
    contribution to the observed chirality arises. In order to
    quantify the toroidal response of the metamaterial, we calculate
    its toroidal moment by the formula,
    $\vec{\tau}=\int{[(\vec{r}\cdot\vec{j})\vec{r}-\vec{r}^2}\vec{j}]d^3r$
    and present in Fig.~4 the corresponding radiated power
    $P_T=\frac{\omega^6\mu_0}{12\pi c^5}|\vec{T}|^2$ \cite{vrejoiu}.
    At most frequencies the toroidal moment is comparable with the
    electric quadrupole moment, which suggests that the toroidal
    contribution to the metamaterial response is in general
    non-negligible. However, as it can be seen from the radiated
    powers, its contribution in the gyrotropy is secondary, since
    the electric and magnetic dipole/electric quadrupole moments are much stronger at
    the dichroism resonances. Nevertheless, a much stronger toroidal
    response is expected at lower frequencies, where electric multipole moments vanish and we intend to study
    this regime in the near future.

\begin{figure}[ht] \label{fig4}
\includegraphics[width=.5\textwidth]
{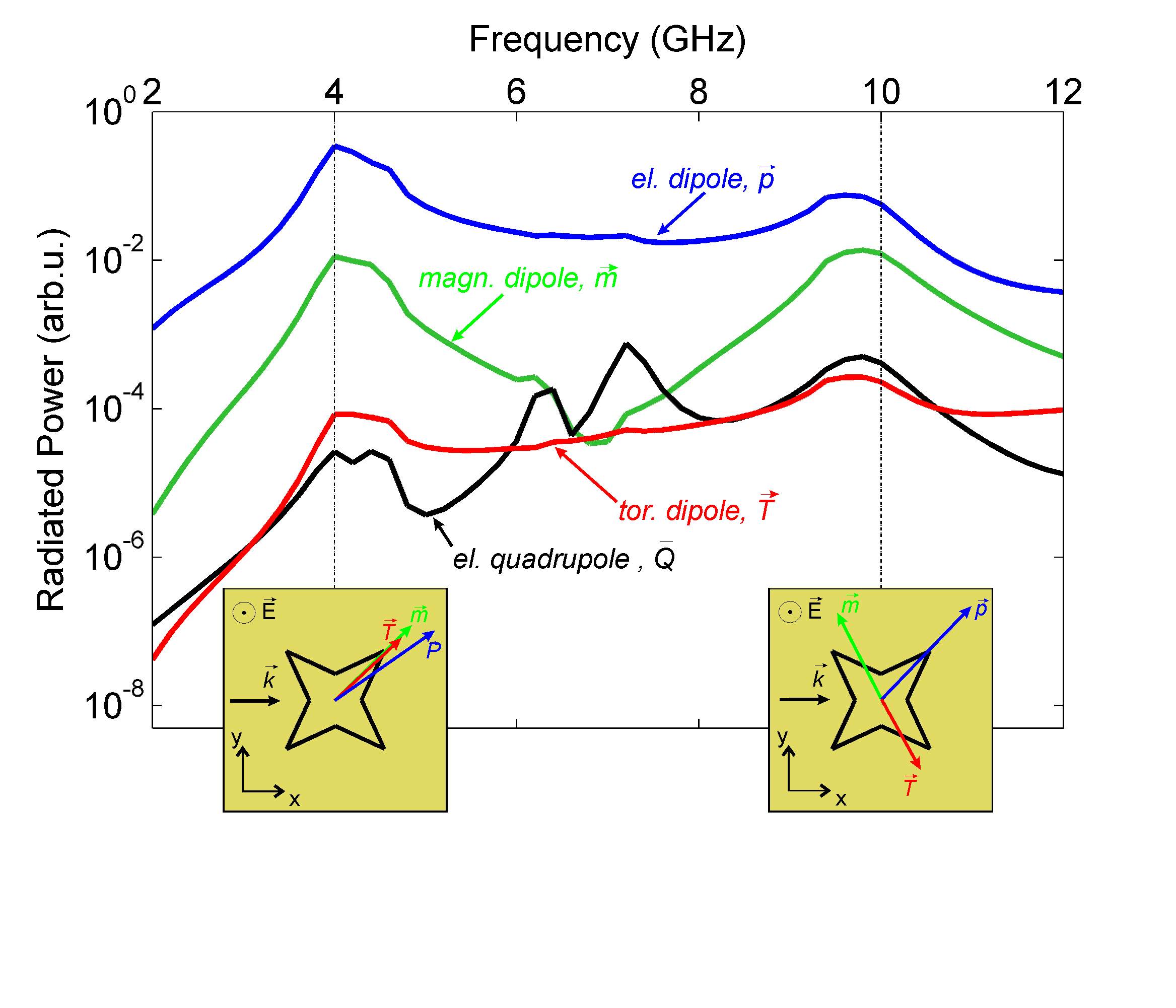}\caption{Radiated power from the different multipole
moments of the (numerically simulated) current configuration on the
wire windings, normalized to the incident power. Insets: Orientation
in the x-y plane of dipole moments at the frequencies marked by the
corresponding dashed lines. In both cases, the z-component of the
moments is much smaller than the x and y components.}
\end{figure}

In conclusion, we present a first experimental investigation of a
new type of chiral metamaterials, which possesses toroidal symmetry.
The studied metamaterial shows strong circular dichroism in the GHz
frequency range as a result of geometrical resonances. The
gyrotropic response of the structure can be explained in terms of
electric and magnetic dipole as well as electric quadrupole moments.
Finally, we discussed the toroidal response of the system and showed
that at specific frequencies it can be of magnetic dipole/electric
quadrupole order.

The authors would like to acknowledge the financial support of the
Engineering and Physical Sciences Research Council, UK.


\newpage


\subsection{References}
\begin{thebibliography}{30}

\bibitem{zeldovich58} Zel'dovich, Ya. B., Zh. Eksp. Teor. Fiz. \textbf{6}, 1184 (1958).
\bibitem{dubovik74} V. M. Dubovik, A. A. Cheshkov, Phys. Part. Nucl. \textbf{5}, 791 (1974).
\bibitem{mainstone67} J. S. Mainstone, Nat. \textbf{215}, 1048 (1967).
\bibitem{dnarev} N. V. Hud and I. D. Vilfan, Annu. Rev. Biophys. Biomol. Struct. \textbf{34}, 295-318 (2005).
\bibitem{protrev} M. M. Hingorani and M. O'Donnell, Nat Rev. Mol. Cell Biol. \textbf{1}, 22-30 (2000).
\bibitem{cyclod} W. Saenger, J. Jacob, K. Gessler, T. Steiner, D. Hoffmann, H. Sanbe, K. Koizumi, S. M. Smith, and T. Takaha, Chem. Rev. \textbf{98}, 1787-1802 (1998).
\bibitem{carbon} J. Lium H. Dai, J. H. Hafner, D. T. Colbert, R. E. Smalley, S. J. Tans, and C. Dekker, Nature \textbf{385}, 780-781 (1997).
\bibitem{ferroelc} A. A. Gorbatsevich and Y. V. Kopaev, Ferroelectrics \textbf{161}, 321-324 (1994).
\bibitem{ferroel} I. I. Naumov, L. Bellaiche and H. X. Fu, Nature \textbf{432}, 737-740 (2004).
\bibitem{afanasiev01} G. N. Afanasiev, J. Phys. D \textbf{34}, 539-559 (2001).
\bibitem{afanasiev98} G. N. Afanasiev and V. M. Dubovik, Phys. Part. Nucl. \textbf{29}, 366 (1998).
\bibitem{sawada05} Sawada K and Nagaosa N 2005 Phys. Rev. Lett. 95 237402
\bibitem{fedotov} V. A. Fedotov, K. Marinov, A. D. Boardman and N. I. Zheludev, New J. Phys. {\bf 9}, 95 (2006).
\bibitem{tormetam} K. Marinov, A. D. Boardman, V. A. Fedotov, and N. I. Zheludev, New J. Phys. {\bf 9}, 324 (2007).
\bibitem{pendry04} J. B. Pendry, Science \textbf{306}, 1353 (2004).
\bibitem{raab} R. E. Raab and O. L. de Lange, \textit{Multipole Theory in Electromagnetism} (Oxford, 2005).
\bibitem{buckingham71} A. D. Buckingham and M. B. Dunn, J. Chemical Soc. A, 1988 (1971).
\bibitem{theron96} I. P. Theron and J. H. Cloete, IEEE Trans. Antennas Propagat. \textbf{44}, 1451 (1996).
\bibitem{vrejoiu} C. Vrejoiu, J. Phys. A: Math. Gen. \textbf{35}, 9911 (2002).
\bibitem{note1} Similar results were obtained under excitation with the orthogonal polarization state (y axis of Fig. 1b).
\bibitem{dubovik90} V. M. Dubovik and V. V. Tugushev, Phys. Rep. \textbf{187}, 145 (1990).
\bibitem{fiebig05} M. Fiebig, J. Phys. D \textbf{38}, R123 (2005).

\end{thebibliography}
\end{document}